\documentclass[doublecol]{epl2}

\title{Mirror symmetry in the energy spectra of $n$-level systems}
\shorttitle{Mirror symmetry in the energy spectra} 

\author{S. Cordero \and O. Casta\~nos \and R. L\'opez--Pe\~na \and E. Nahmad--Achar}

\institute{%
Instituto de Ciencias Nucleares, Universidad Nacional Aut\'onoma de M\'exico, Apartado Postal 70-543, 04510 M\'exico DF,   Mexico  }


\abstract{
The energy spectrum of a system of $N_a$ atoms of $n$ levels interacting with a one-mode electromagnetic field is studied in the dipole and rotating wave approximations.  We find that, under the resonant condition, it exhibits a mirror symmetry with respect to the energy $E=M$ where $M$ the total number of excitations.  Thus, for any eigenstate  $|\psi_M^{+}\rangle$ with energy $E=M+{\cal E}$ there exists a related eigenstate  $|\psi_M^{-}\rangle$ with energy $E=M-{\cal E}$ via the unitary parity operator in the number of photons .  This is independent of the dipolar  coupling between  the levels. We give explicit examples for $3$-level systems.  
}

\begin{document}
\maketitle

\section{ Introduction} 

Systems describing the interaction of a collection  of $N_a$ atoms of $n$-levels  with a one-mode quantized electromagnetic field in the dipolar and rotating wave approximations have an extensive use in quantum optics~\cite{dodonov03}. For $n=2$, the Tavis-Cummings model~\cite{tavis68} has been physically realized using QED cavity with Bose-Einstein condensates \cite{baumann10,nagy10}. The description of phase transitions in two-level systems has been studied in~\cite{castanos09a,castanos09b,nahmad-achar13}.

The dynamical behavior of one atom with two or three levels interacting with quantized cavity fields was reviewed in~\cite{yoo85}.
The Hamiltonian describing a three-level ladder atom interacting with a broadband squeezed vacuum field was used to derive the master equation for a reduced density operator of the atom in~\cite{ficek91},
while phase operators to describe three-level systems in the $\Lambda$-configuration of one atom, to deal with quantum interference effects in atom-field interactions, were introduced in~\cite{klimov03}.
The time evolution of the second-order correlation function when a three-level atom interacting with a single mode cavity field is initially in an upper state and the field is in a coherent state, was studied in~\cite{abdel-wahab07}.

The energy surface method was used in~\cite{cordero1} to determine the phase diagram in the dipolar strength space for a finite number of three-level atoms interacting with a one-mode radiation field in all three configurations, and a comparison with symmetry-adapted projected states was made in~\cite{cordero2}, together with a study of the statistics of the total number of excitations and the number of photons.
Three-level systems have also been proposed for quantum memories and quantum logical gates~\cite{fleischhauer02,yi03,jane02}.

In this paper we show that for a collection of $N_a$ atoms of $n$-levels interacting with an electromagnetic field, in the dipolar and rotating wave approximations (RWA), the {\em resonant} condition ({\em i.e.}, a system with zero detuning) adds a {\em mirror} symmetry to the energy spectrum. The corresponding symmetric (reflected) eigenstates are related via a {\em photon parity} operator.  Hence, one may only distinguish these states via physical quantities that anti-commute with this photon parity operator.

\section{ Model} 
Consider the Hamiltonian for $N_a$ atoms of $n$-levels interacting in the RWA and dipolar approximation with a one-mode electromagnetic field,
\begin{equation}\label{eq.H.1}
{\bm H} = {\bm H}_D+{\bm H}_{int}, 
\end{equation}
where ${\bm H}_D$ and ${\bm H}_{int}$ are, respectively, the diagonal and interaction contributions ($\hbar=1$)~\cite{cordero2}
\begin{eqnarray}
{\bm H}_D &=& \Omega \,{\bm a}^\dag\,{\bm a} + \sum_{k=1}^n\omega_k\, {\bm A}_{kk}, \label{eq.HD} \\[3mm]
{\bm H}_{int}&=& -\frac{1}{\sqrt{N_a}}\sum_{k<l} \mu_{kl}\left({\bm a}^\dag \,{\bm A}_{kl} + {\bm a}\,{\bm A}_{lk}\right).\label{eq.Hint}
\end{eqnarray}
Here ${\bm a}^\dag,\,{\bm a}$ are the usual creation and annihilation field operators, and ${\bm A}_{ij}$ the matter operators obeying the $U(n)$ algebra
\begin{equation}
\left[ {\bm A}_{ij},{\bm A}_{lm}\right] = \delta_{jl}\,{\bm A}_{im}-\delta_{im}\,{\bm A}_{lj}.
\end{equation}
The total number of atoms is given by
\begin{equation}\label{eq.Na}
N_a=\sum_{k=1}^n {\bm A}_{kk}.
\end{equation}
The parameters $\Omega$ and $\omega_k$ are, respectively, the frequencies of the field and atomic energy levels,  and  $\mu_{ij}$ the intensity of the dipolar coupling (control parameters) between levels $i$ and $j$; we consider $\mu_{ji}=\mu_{ij}$.

Without loss of generality, we fix the  lowest atomic energy level at $\omega_1=0$ and use the labeling  $\omega_i\leq\omega_j$ for $i<j$. The different atomic configurations are chosen by  taking the appropriate value $\mu_{ij}=0$ indicating that the transition $i\leftrightarrow j$ is forbidden.    

Besides the total number of atoms, there exists for each atomic configuration an additional constant of motion, namely, the {\em total number of excitations}
\begin{eqnarray}\label{eq.M}
{\bm M} = {\bm n} + \sum_{k=2}^n\lambda_k\,{\bm A}_{kk},
\end{eqnarray}
where ${\bm n}={\bm a}^\dag\,{\bm a}$ is the photon number operator and the integer values $0\leq\lambda_k\leq n-1$ depend on the corresponding atomic configuration. The values $\lambda_k$ may be interpreted as the number of photons that are required to excite one atom from the lowest level $\omega_1$ to the $k$-excited level $\omega_k$.

\section{ Mirror symmetry}
Substituting  Eqs.~(\ref{eq.Na}) and (\ref{eq.M}) into the diagonal contribution of the Hamiltonian ${\bm H}_D$ one has ($\omega_1=0$)
\begin{eqnarray}
{\bm H}_D &=& \Omega \,{\bm M}  + \sum_{k=2}^n \left(\omega_k-\lambda_k\,\Omega\right){\bm A}_{kk}.
\end{eqnarray}
For a given atomic configuration, the {\em resonant} condition $\omega_k-\lambda_k\,\Omega=0$ may be obtained simultaneously for all $k$ values, and when this is so the diagonal contribution ${\bm H}_D=\Omega\,{\bm M}$ is a  constant of motion. So, under the resonant condition, the eigenvalues of the Hamiltonian~(\ref{eq.H.1}) take the form $E=\Omega\, M  + {\cal E}_{int}$, for a fixed $M$ value,  where ${\cal E}_{int}$ is the contribution of the interaction term~(\ref{eq.Hint}). 

We now consider the {\em photon parity} operator defined as
\begin{equation}\label{eq.P}
{\bm P} = e^{i\,\pi\, {\bm n}}. 
\end{equation}
Clearly, ${\bm P}$ commutes with ${\bm H}_D$, but it anti-commutes with the interaction term: ${\bm H}_{int}\,{\bm P} = -{\bm P}\,{\bm H}_{int}$. So, for an eigenstate $|\psi_M\rangle$ satisfying  
\begin{equation}
{\bm H}\,|\psi_M\rangle = \left(\Omega\, M + {\cal E}_{int}\right)|\psi_M\rangle,
\end{equation}
the state $|\psi_M'\rangle$ defined as $|\psi_M'\rangle = {\bm P}|\psi_M\rangle$ satisfies  
\begin{equation}
{\bm H}\,|\psi_M'\rangle = \left(\Omega\, M  - {\cal E}_{int}\right)|\psi_M'\rangle.
\end{equation}

The previous result shows that there is a mirror symmetry in the energy spectrum around $E=\Omega\, M $, independently of the dipolar couplings between the levels. Additionally, the eigenstate $|\psi_M^-\rangle$ and its corresponding {\em reflected} eigenstate $|\psi_M^+\rangle={\bm P}\,|\psi_M^-\rangle$ obey the same statistics (they have the same expectation values of the diagonal operators, such as number of photons and atomic populations, including their corresponding fluctuations), since the operator ${\bm P}$ only changes the state component phases according to the parity of the number of photons, i.e., changes only {\it local phases}. 

Furthermore, states $|\psi_M\rangle$ for which $E=\Omega\,M$ (when they exist) are their self mirror image, as ${\bm H}_{int}|\psi_M\rangle = 0$ is necessarily satisfied. For these particular states one finds ${\bm P}|\psi_M\rangle = \pm|\psi_M\rangle$, {\em i.e.}, these states have only even or odd contributions of the number of photons and hence ${\bm P}$ plays strictly the role of a parity operator: if there is degeneracy, all states with energy $E_M=\Omega\,M$ possess the same parity of the photon contribution.

From the above one may conclude that,  if $|\psi_M\rangle$ is an eigenstate of ${\bm H}$  and the resonant condition is satisfied, then  $|\psi_M\rangle$ has energy $E=\Omega\,M$ if, and only if, $|\langle  \psi_M|{\bm P}|\psi_M\rangle|^2 =1$. 

The  mirror energy symmetry of the expectation value of the Hamiltonian holds also for an arbitrary state $|\Psi\rangle$. In this case, if $\langle {\bm H}\rangle_{|\Psi\rangle} = E_D +E_{int}$ (where  $E_D$ and $E_{int}$ stand, respectively, for the expectation values of the diagonal and interaction terms), then the state ${\bm P}|\Psi\rangle$ satisfies  $\langle {\bm H}\rangle_{{\bm P}|\Psi\rangle} = E_D -E_{int}$. However, when $E_{int}=0$ the state $|\Psi\rangle$ is expanded only into eigenstates with energy $E=\Omega\,M$.

\section{Example} 
In order to exemplify our result, we consider $N_a$ atoms of three levels interacting with a one-mode electromagnetic field in the resonant condition. For this system there are three different atomic configurations, namely, $\Xi,\,\Lambda$ and $V$, with total number of excitations operator ${\bm M}_\Xi = {\bm n} + {\bm A}_{22} + 2\,{\bm A}_{33},\ {\bm M}_\Lambda = {\bm n} + {\bm A}_{33}$ and ${\bm M}_V = {\bm n} + {\bm A}_{22} + {\bm A}_{33}$, respectively.

\begin{table}
\begin{center}
\begin{tabular}{c|c |c| c}
&$\Xi\, (\mu_{13}=0)$ & $ \Lambda\, (\mu_{12}=0)$ & $ V\, (\mu_{23}=0)$\\ \hline &&& \\[-2mm]
$\omega_2$ &1 & 0& 1\\[1mm]
$\omega_3$ &2 & 1& 1\\ \hline
\end{tabular}
\caption{Frequency values for the atomic levels in each configuration, under the resonant condition. We use $\omega_{1}= 0$ throughout.}\label{t1}
\end{center}
\end{table}

In a recent work~\cite{castanosQTS8} it was pointed out that the dimension of the Hilbert space of this kind of system is strongly dependent on the values of $M$ and $N_a$, except for large values of $M$, in which  case the dimension is given by $(N_a+1)(N_a+2)/2$. A similar situation is found when we consider the number of eigenstates with energy $E=\Omega\,M$: denoting by $\lfloor\,x\,\rfloor$ the floor value of $x$, the number of eigenstates with energy $E=\Omega\,M$ supported by the Hamiltonian is given by\\
$\Xi$-configuration:
\begin{eqnarray}\label{eq.deg.X}
\hbox{$M$ even }&&\left\{\begin{array}{ll}
\frac{M}{2}+1\ ,& M\leq N_{a}\\[3mm]
\lfloor\frac{N_{a}}{2}\rfloor+1\ ,& M> N_{a}
\end{array}\right. \nonumber \\
&\\
\hbox{$M$ odd }&&\left\{\begin{array}{ll}
0\ ,& M\leq N_{a}\\[1mm]
\lfloor\frac{M-N_{a}}{2}\rfloor+\frac{1+(-1)^{N_{a}}}{2}\ , &N_{a}
< M<2\,N_{a}\\[2mm]
\lfloor\frac{N_{a}}{2}\rfloor+1\ ,& M> 2\,N_{a}
\end{array}\right. \nonumber
\end{eqnarray}  
$\Lambda$-configuration:
\begin{eqnarray}\label{eq.deg.L}
\hbox{$M$ even }&&\left\{\begin{array}{ll}
N_a-\frac{M}{2}+1\ ,& M\leq N_{a}\\[3mm]
\lfloor\frac{N_{a}}{2}\rfloor+1\ ,& M> N_{a}
\end{array}\right. \nonumber \\
&\\
\hbox{$M$ odd }&&\left\{\begin{array}{ll}
\lfloor\frac{M}{2}\rfloor+1 & M\leq N_{a}\\[2mm]
\lfloor\frac{N_{a}}{2}\rfloor+1\ ,& M> N_{a}
\end{array}\right. \nonumber
\end{eqnarray}  
$V$-configuration: 
\begin{eqnarray}\label{eq.deg.V}
M\hbox{ even or odd }
\left\{\begin{array}{ll}
\lfloor\frac{M}{2}\rfloor+1\ ,& M \leq N_{a}\\[3mm]
\lfloor\frac{N_{a}}{2}\rfloor+1\ ,& M > N_{a}
\end{array}\right.
\end{eqnarray}
Notice that, in all cases, for large values of $M$ ($\geq 2\,N_a$ for the $\Xi$-configuration, and $\geq N_a$ for the $\Lambda$- and $V$-configurations) the Hamiltonian supports $\lfloor N_{a}/2\rfloor+1$ states with energy $E=\Omega\,M$.

As a numerical example, the three atomic configurations will now be considered for $N_a=10$ atoms. We choose $\Omega=1,\,\omega_1=0$, and the atomic levels in resonant condition given in table \ref{t1}.

\begin{center}
\begin{figure}[t!]
\includegraphics[width=0.9\linewidth]{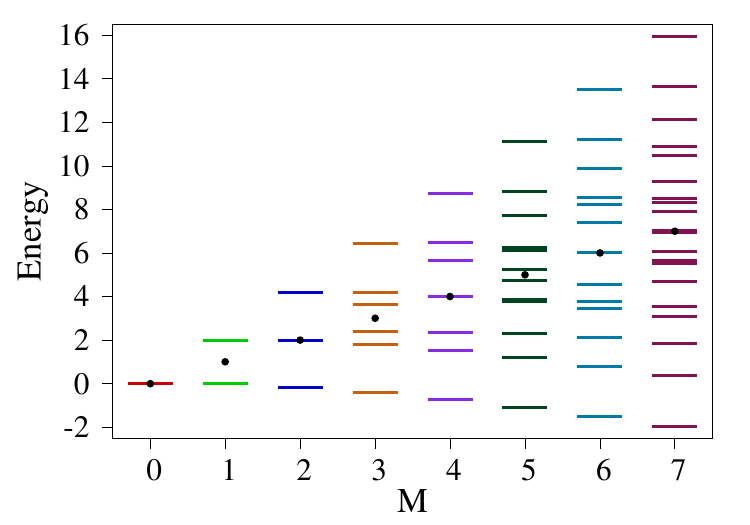}
\caption{(Color online) Energy spectra for different values of the total number of excitations $M$, for atoms in the $\Xi$-configuration. The values $E=M$ are indicated by dots. The parameters are $\mu_{12}=1$ and $\mu_{23}=3$.}\label{f1}
\end{figure}
\end{center}

Fig.~\ref{f1} shows, for atoms in the $\Xi$-configuration, the energy spectra for different values of the total number of excitations $M$. One may see that only the first even values of $M$ provide eigenstates with energy $E=M$, in accordance with Eqs.~(\ref{eq.deg.X}). The values $E=M$ are indicated by dots, in order to appreciate visually that the mirror symmetry  appears around these values. A similar behavior of the energy spectra is obtained when the other configurations are considered, but both the $\Lambda$- and $V$-configurations always support eigenstates with energy $E=M$ (Eqs.~(\ref{eq.deg.L}) and (\ref{eq.deg.V})).

\begin{center}
\begin{figure}
\includegraphics[width=0.9\linewidth]{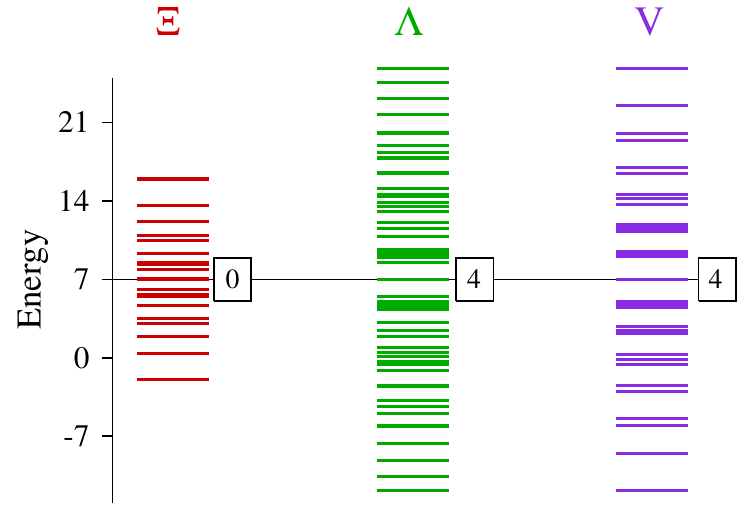}
\caption{(Color online)  Energy spectra of the different configurations of $3$-level atoms for $M=7$ excitations. The number of eigenstates with $E=M$ is shown. The parameters are $(\mu_{12},\mu_{23}) =(1,3)$ for the $\Xi$-configuration, $(\mu_{23},\mu_{13}) =(1,3)$ for the $\Lambda$-configuration, and $(\mu_{12},\mu_{13}) =(1,3)$ for the $V$-configuration. }\label{f2}
\end{figure}
\end{center}

Fig.~\ref{f2} shows the energy spectra of the different atomic configurations $\Xi,\,\Lambda$ and $V$, for a fixed number of excitations $M=7$, exhibiting the mirror symmetry with respect to the energy $E=7$. The number of states with energy $E=M$ is shown. Note that the ground state for the $\Lambda$- and $V$-configurations have the same energy value, under the resonant condition.  This is also true for any value of the total number of excitations.

\begin{center}
\begin{figure}[t!]
\includegraphics[width=0.9\linewidth]{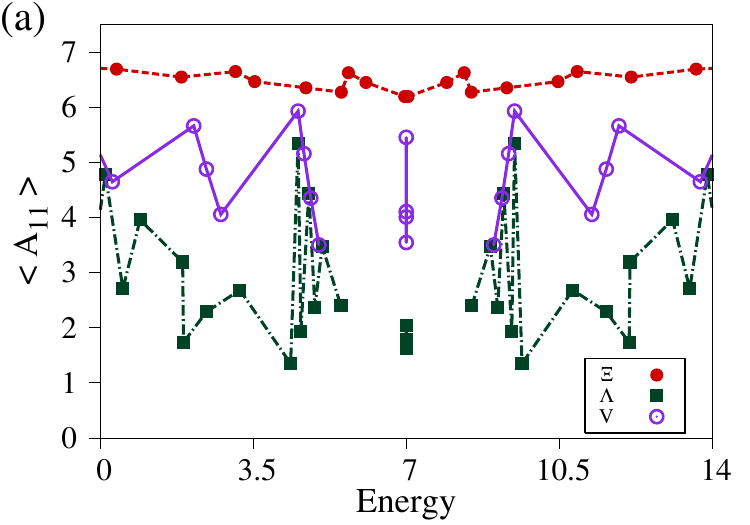}\\[2mm]
\includegraphics[width=0.9\linewidth]{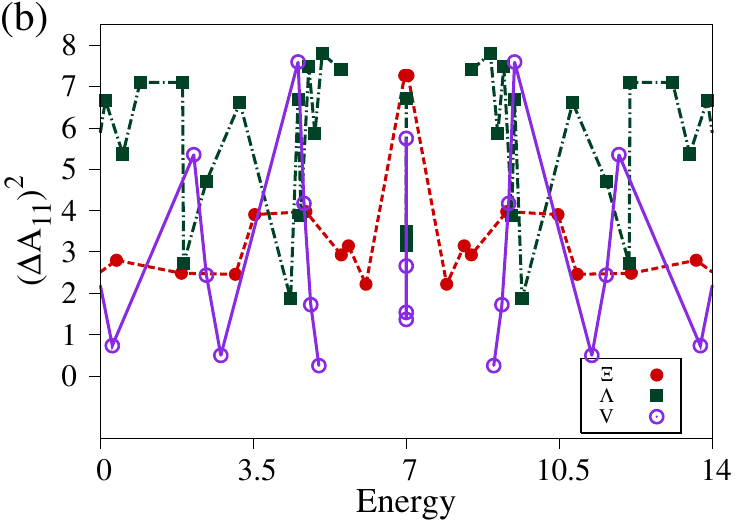}\\[2mm]
\includegraphics[width=0.9\linewidth]{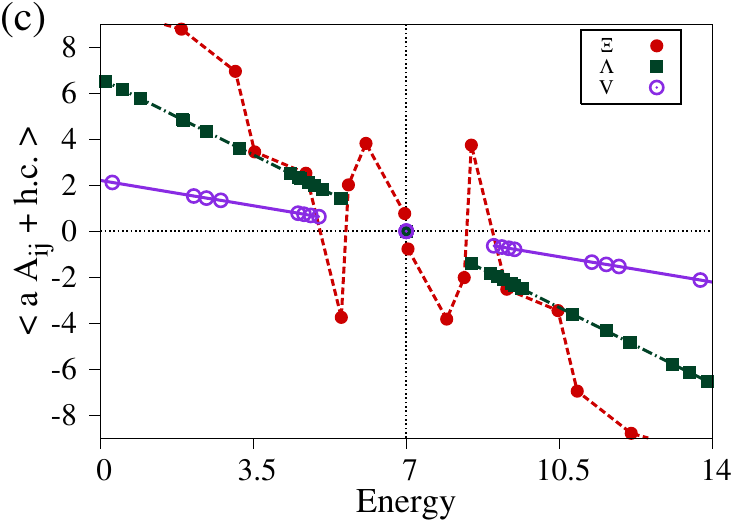}
\caption{(Color online)  (a) Expectation value of the ${\bm A}_{11}$, (b) its fluctuation $(\Delta A_{11})^2$, and (c) the expectation value of ${\bm a}\,{\bm A}_{ij}+h.c.$ with $i=2,\,j=1$ for $\Xi$- and $V$-configurations and $i=3,\,j=1$ for $\Lambda$-configuration. Parameters as in Fig.~\ref{f2}. }\label{f3}
\end{figure}
\end{center}

As pointed out above, the eigenstates $|\psi_M^\pm\rangle$ obey the same statistics since they differ only by local phases and hence one cannot distinguish them using observables that commutes with ${\bm P}$. As an example of this fact, we calculate the expectation value of the atomic population of the  lowest atomic level, ${\bm A}_{11}$, with respect to the eigenstates of the system.  Fig.~\ref{f3}(a) shows that expectation value, $\langle {\bm A}_{11}\rangle$,  as function of the energy for the different atomic configurations ($\Xi$ solid circles, $\Lambda$ solid squares and $V$ empty circles). All of them exhibit the mirror symmetry around $E=7$. A similar situation occurs for its corresponding fluctuation $(\Delta A_{11})^2$ as it is shown in Fig.~\ref{f3}(b). In order to distinguish the states $|\psi^\pm_M\rangle$ we consider an operator that does not commute with ${\bm P}$.  One may consider an operator of the form ${\bm a}{\bm A}_{ij}+{\bm a}^\dag{\bm A}_{ji}$, which cannot change the total number of excitations, $M$.  For the $\Xi$- and $V$-configurations we choose for the operator   $i=2,\, j=1$  while for the $\Lambda$-configuration we use $i=3,\,j=1$. Fig.~\ref{f3}(c) shows the expectation value of the operator $\langle {\bm a}{\bm A}_{ij}+{\bm a}^\dag{\bm A}_{ji}\rangle$ with respect to the eigenstates as a function of the energy. 
Note the change of sign with respect to $E=M$, allowing us to distinguish between the states $|\psi\rangle$ and  ${\bm P}|\psi\rangle$.

\begin{center}
\begin{figure}[t!]
\includegraphics[width=0.9\linewidth]{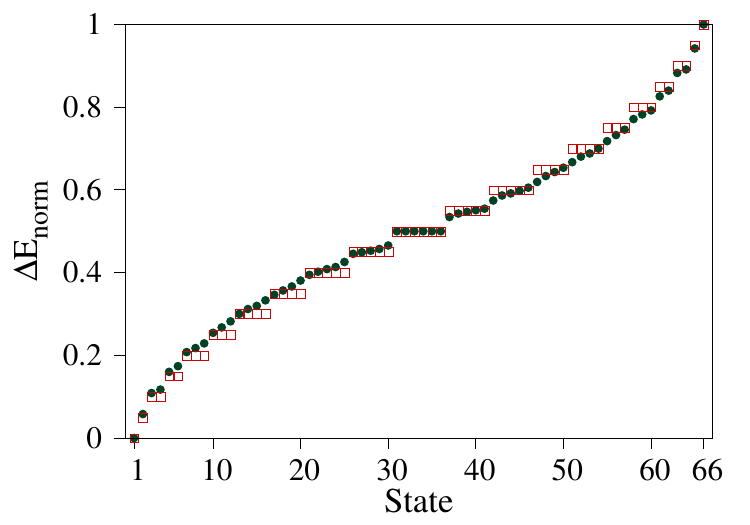}
\caption{(Color online) Renormalized energy spectrum in the $\Xi$-configuration of $N_a=10$ atoms and  $M=20$ (solid dots) compared with the limit $M\to \infty$ (empty squares). We use $\mu_{12}=1$ and $\mu_{23}=\sqrt 2$.}\label{f4}
\end{figure}
\end{center}

The limit $M\to\infty$ corresponds to a classical field in the Hamiltonian (\ref{eq.H.1}). In order to compare the energy spectrum with a finite number of excitations and its semi-classical limit, we renormalize the energy as  
\begin{equation}
\Delta E_{norm} := \frac{E - E_0}{E_{max}-E_0},
\end{equation}
where $E_{max}$ and $E_0$ stand for the highest and lowest energies, respectively. Fig.~\ref{f4} shows the renormalized energy spectrum for a system of $N_a=10$ atoms in the $\Xi$-configuration with $M=20$ (solid dots). This is compared with the classical limit $M\to \infty$ (empty squares). Both exhibit the same qualitative behavior. The mirror symmetry remains in the limit $M\to\infty$, although now the spectra is also degenerate in levels with $E\neq M$. 
%

\section{ In summary}
We consider a system of $N_a$ atoms of $n$-levels interacting via RWA and dipolar approximation with a one-mode quantized electromagnetic field.  We find that, under a resonant condition, this kind of systems present a mirror energy symmetry around the value $E=M$, independently of the intensity of the dipolar coupling constants (see Figs.~\ref{f1} and \ref{f2} for the particular case of $3$-level atoms).  The reflected eigenstates $|\psi^\pm_M\rangle$ with eigenvalues $E=M\pm{\cal E}$ are related via the photon parity operator ${\bm P}$;  hence, these states possess the same expectation values of physical quantities that commutes with ${\bm P}$, including their fluctuations (see Figs.~\ref{f3}(a) and \ref{f3}(b)).  On the other hand, when a physical quantity does not commute with ${\bm P}$, the expectation value only differs by  a sign [see Fig.~\ref{f3}(c)].

For $N_a$ atoms of $3$-levels interacting with a one-mode electromagnetic field in the resonant condition,  we find that the number of states with $E=M$ strongly depends on the values of $M$ and $N_a$ (cf. Eqs.~(\ref{eq.deg.X})--(\ref{eq.deg.V})), except for large values of $M$, where the number of states with $E=M$ is given by $\lfloor N_a/2\rfloor+1$ independently of the atomic configuration. 

These results may be generalized to $n$-level atoms interacting with two or more modes of an electromagnetic field, under similar considerations, since the mirror symmetry of the energy spectrum appears due to the fact that the diagonal contribution ${\bm H}_D$ is a  constant of motion. \\

\acknowledgments
This work was partially supported by CONACyT-M\'exico (under project 101541), and DGAPA-UNAM (under projects IN101614 and IN110114).

\end{document}